\documentclass[aps,prb,twocolumn,superscriptaddress,showpacs]{revtex4}

\usepackage{graphicx}
\usepackage{dcolumn}
\usepackage{bm}

\begin{document}

\title{Low-temperature heat transport of the geometrically
frustrated antiferromagnets $R_2$Ti$_2$O$_7$ ($R$ = Gd and Er)}

\author{F. B. Zhang}
\affiliation{Hefei National Laboratory for Physical Sciences at
Microscale, University of Science and Technology of China, Hefei,
Anhui 230026, People's Republic of China}

\author{Q. J. Li}
\affiliation{Hefei National Laboratory for Physical Sciences at
Microscale, University of Science and Technology of China, Hefei,
Anhui 230026, People's Republic of China} \affiliation{School of
Physics and Material Sciences, Anhui University, Hefei, Anhui
230601, People's Republic of China}

\author{Z. Y. Zhao}
\affiliation{Hefei National Laboratory for Physical Sciences at
Microscale, University of Science and Technology of China, Hefei,
Anhui 230026, People's Republic of China}

\author{C. Fan}
\affiliation{Hefei National Laboratory for Physical Sciences at
Microscale, University of Science and Technology of China, Hefei,
Anhui 230026, People's Republic of China}

\author{S. J. Li}
\affiliation{Hefei National Laboratory for Physical Sciences at
Microscale, University of Science and Technology of China, Hefei,
Anhui 230026, People's Republic of China}

\author{X. G. Liu}
\affiliation{Hefei National Laboratory for Physical Sciences at
Microscale, University of Science and Technology of China, Hefei,
Anhui 230026, People's Republic of China}

\author{X. Zhao}
\email{xiazhao@ustc.edu.cn} \affiliation{School of Physical
Sciences, University of Science and Technology of China, Hefei,
Anhui 230026, People's Republic of China}

\author{X. F. Sun}
\email{xfsun@ustc.edu.cn} \affiliation{Hefei National Laboratory
for Physical Sciences at Microscale, University of Science and
Technology of China, Hefei, Anhui 230026, People's Republic of
China}

\date{\today}

\begin{abstract}

We report a systematic study on the low-temperature thermal
conductivity ($\kappa$) of $R_2$Ti$_2$O$_7$ ($R$ = Gd and Er)
single crystals with different directions of magnetic field and
heat current. It is found that the magnetic excitations mainly act
as phonon scatterers rather than heat carriers, although these two
materials have long-range magnetic orders at low temperatures. The
low-$T$ $\kappa(H)$ isotherms of both compounds show rather
complicated behaviors and have good correspondences with the
magnetic transitions, where the $\kappa(H)$ curves show drastic
dip- or step-like changes. In comparison, the field dependencies
of $\kappa$ are more complicated in Gd$_2$Ti$_2$O$_7$, due to the
complexity of its low-$T$ phase diagram and field-induced magnetic
transitions. These results demonstrate the significant coupling
between spins and phonons in these materials and the ability of
heat-transport properties probing the magnetic transitions.

\end{abstract}


\maketitle

\section{INTRODUCTION}

Geometrically frustrated antiferromagnets have been a topic of
significant interest due to their abundant and exotic physical
properties.\cite{Schiffer, Lecheminant, Mila, Canals, Gardner} The
rare-earth titanates $R_2$Ti$_2$O$_7$ ($R =$ rare earth) are one
of the families with three-dimensional spin
frustration.\cite{Canals, Gardner} These materials have pyrochlore
crystal structure with the space group $Fd\bar{3}m$, in which the
magnetic rare-earth ions form a network of corner-sharing
tetrahedra and are prone to a high degree of geometric
frustration. For this reason, antiferromagnetically coupled
classical Heisenberg spins on the pyrochlore lattice cannot form a
long-range order at finite temperature. However, the real
materials are greatly sensitive to weak perturbations (e.g.
single-ion anisotropy, dipolar interaction or quantum
fluctuations) beyond the nearest-neighboring exchange, which
results in rich unconventional low-temperature magnetic and
thermodynamic properties. Example phenomena include the disordered
classical spin ice in Ho$_2$Ti$_2$O$_7$ and Dy$_2$Ti$_2$O$_7$
(with effective ferromagnetic exchange and Ising
anisotropy),\cite{Bramwell_HTO, Ramirez_DTO} the quantum spin
liquid in Tb$_2$Ti$_2$O$_7$,\cite{Gardner_TTO1, Gardner_TTO2,
Yin_TTO} and the quantum spin ice in
Yb$_2$Ti$_2$O$_7$.\cite{Shannon_YbTO, Applegate_YbTO} Furthermore,
the long-range antiferromagnetic (AF) order can be formed in
Gd$_2$Ti$_2$O$_7$ and Er$_2$Ti$_2$O$_7$.\cite{Raju, Champion1}
Recently, the low-$T$ heat transport properties of
$R_2$Ti$_2$O$_7$ have started to attract research interests for
the purpose of detecting the transport of magnetic excitations in
these materials. The spin-ice compounds exhibit considerably
strong interactions between the magnetic excitations and phonons
and hence show drastic changes of thermal conductivity at the
field-induced magnetic transitions.\cite{Fan_DTO, Kolland, Toews}
In the spin-liquid material Tb$_2$Ti$_2$O$_7$, the coupling
between spin fluctuations and phonons is so strong that the phonon
heat transport is extremely weak.\cite{Li_TTO} In this work, we
study the low-$T$ thermal conductivity of two long-range ordered
systems $R_2$Ti$_2$O$_7$ ($R =$ Gd and Er) to probe the roles of
magnon excitations and magnetic fluctuations.

In Gd$_2$Ti$_2$O$_7$, the magnetic Gd$^{3+}$ ion has a spin
momentum $S =$ 7/2 and an orbital momentum $L =$ 0, and therefore
has a negligible single-ion anisotropy.\cite{Raju} In zero field,
Gd$_2$Ti$_2$O$_7$ shows the magnetic-order transition induced by
thermal fluctuation at about 1 K to the so-called $P$
state.\cite{Stewart, Champion2, Hassan, Reimers, Enjalran}
Lowering temperature to about 0.7 K, the $P$ state changes to a
so-called $F$ state. It was found that both $P$ and $F$ states
have a specific 4-$k$ magnetic structure with {\bf k} = (1/2, 1/2,
1/2), in which all the Gd$^{3+}$ spins are perpendicular to the
local [111] axes.\cite{Stewart} However, in the $F$ state each
pyrochlore unit cell consists of three fully ordered spin
tetrahedra and one weakly ordered spin tetrahedron, while in the
$P$ state that weakly ordered spin tetrahedron becomes fully
disordered.\cite{Stewart} With applying magnetic field, the
Gd$^{3+}$ spin structures present further changes.\cite{Ramirez,
Petrenko1, Glazkov1, Sosin1, Petrenko2} At $H \geq$ 3 T and $T <$
1 K, a collinear spin state is formed, in which three spins on
each tetrahedron are aligned with the field and the rest one
points to the opposite direction.\cite{Petrenko1} In high magnetic
fields, the Gd$^{3+}$ spins enter a polarized paramagnetic state.
The mechanisms of these field-driven transitions are likely that
magnetic field can break the degenerate ground states and force
the system to select a particular ordered structure.

Er$_2$Ti$_2$O$_7$ was proposed to be a realization of the XY
antiferromagnet on pyrochlore lattice. Er$^{3+}$ ion has a large
orbital momentum $L$ = 6 and a spin momentum $S$ = 3/2. However,
the crystal-field ground state of Er$^{3+}$ ion is a Kramers
doublet, with large anisotropy respective to the local [111] axis,
and can be described as an effective spin of 1/2.\cite{Champion1,
Dalmas} This implies a significant quantum effect in this
material. Er$_2$Ti$_2$O$_7$ undergoes a second-order transition to
a long-range ordered state at about 1.2 K with an unusual {\bf k}
= 0 non-coplanar AF structure, the so-called $\psi_2$
state.\cite{Champion1, Ruff, Sosin2, Dalmas, Zhitomirsky, Savary,
Poole, Cao1} The mechanism of this long-range ordered phenomenon
had been discussed to be the quantum order by disorder
effect,\cite{Bramwell, Zhitomirsky} for which the quantum
fluctuations lift the degeneracy of the ground state. Applying
magnetic field above a critical value of 1.5--2 T (slightly
different for $H \parallel$ [100], [110], and [111]), the system
enters a high-field quantum paramagnetic state through a
second-order transition driven by the quantum fluctuations. In the
high-field state, the Er$^{3+}$ spins are fully polarized in the
XY plane and maximize their projections along the field
direction.\cite{Ruff, Cao1, Cao2}

Both Gd$_2$Ti$_2$O$_7$ and Er$_2$Ti$_2$O$_7$ exhibit the ground
states of a coexistence of the long-range order and short-range
order or spin fluctuations,\cite{Champion1, Ruff, Dunsiger,
Yaouanc} which are obviously different from the conventional
magnetic materials. One can naturally expect that their heat
transport properties would show some peculiar phenomena. In this
paper, we perform detailed studies of the low-$T$ thermal
conductivity of $R_2$Ti$_2$O$_7$ ($R$ = Gd and Er) single crystals
and find that they have rather complicated temperature and
magnetic-field dependencies. To probe the mechanism of the heat
transport properties, the low-$T$ specific heat are also studied.
The data demonstrate that the magnetic excitations in these
materials can effectively scatter phonons instead of transport
heat. Due to the spin-phonon coupling, the field-induced phase
transitions cause drastic changes of the phonon thermal
conductivity.

\section{EXPERIMENTS}

High-quality $R_2$Ti$_2$O$_7$ ($R$ = Gd and Er) single crystals
were grown using a floating-zone technique.\cite{Li_Growth} In
particular, Gd$_2$Ti$_2$O$_7$ and Er$_2$Ti$_2$O$_7$ were grown in
0.4 MPa pure oxygen with a growth rate of 2.5 mm/h and in 0.1 MPa
pure oxygen with a rate of 4 mm/h, respectively. The crystals were
oriented by the x-ray Laue photograph and cut precisely along
those crystallographic axes like [111], [110] (or [1$\bar{1}$0]),
and [11$\bar{2}$] into long-bar and thin-plate shapes for the
thermal conductivity and specific heat measurements, respectively.
The thermal conductivity was measured at low temperatures down to
0.3 K and in magnetic fields up to 14 T by using a conventional
steady-state technique.\cite{Fan_DTO, Li_TTO, Sun_DTN, Zhao_GFO,
Wang_HMO, Wang_TMO, Li_NGSO} The specific heat was measured at low
temperatures down to 0.4 K by using the relaxation method in a
commercial physical property measurement system (PPMS, Quantum
Design).

\section{RESULTS}

\subsection{Gd$_2$Ti$_2$O$_7$}

\begin{figure}
\includegraphics[clip,width=8.5cm]{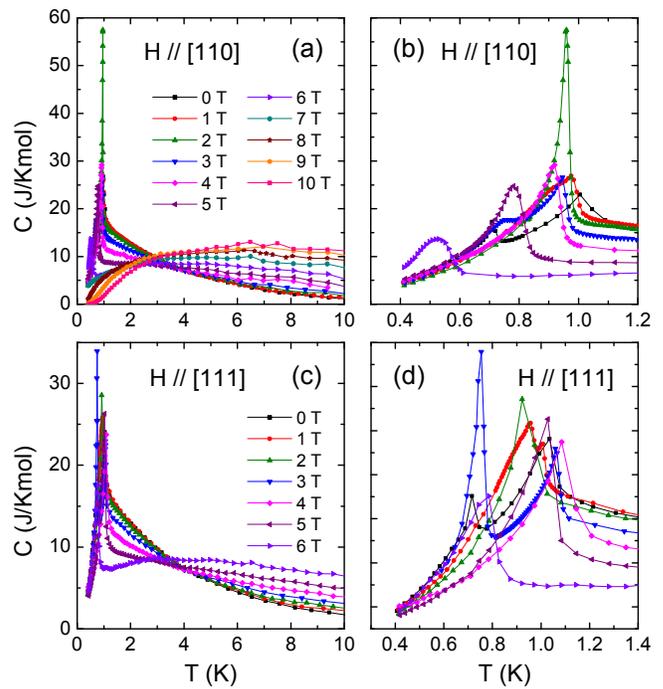}
\caption{(Color online) (a,c) Temperature dependencies of the
specific heat of Gd$_2$Ti$_2$O$_7$ single crystals for $H
\parallel$ [110] and $H \parallel$ [111]. (b,d) Zoom-in of the
low-temperature plots. The sample sizes are
0.29$\times$0.32$\times$0.17 mm$^{3}$ for $H \parallel$ [110] and
0.53$\times$0.53$\times$0.09 mm$^{3}$ for $H \parallel$ [111],
respectively. The magnetic fields were applied along the shortest
dimensions of the samples.}
\end{figure}

Figure 1 shows the low-$T$ specific heat data of Gd$_2$Ti$_2$O$_7$
single crystals for $H \parallel$ [110] and $H \parallel$ [111].
In zero magnetic field, there are two peaks at 1.0 and 0.71 K,
which represent two consecutive transitions from paramagnetic
state to the $P$ state and then to the $F$ state, respectively. In
applied fields along the [110] direction, the two peaks merge with
each other and a broad peak can be seen at 1 T. Increasing field
to 2 T, the peak becomes much larger and a bit sharper and is
locating at 0.96 K. However, this peak splits into a peak at 0.95
K and a lower shoulder at 0.77 K when magnetic field is increased
to 3 T. Increasing field further to 4 T, the peak shifts to 0.92 K
while the shoulder-like feature disappears. For $H >$ 4 T, the
remained peak gradually moves to lower temperatures and cannot be
seen at temperature down to 0.4 K when the field is larger than 6
T. Finally, a hump-like feature appears and gradually shifts to
high temperatures for $H >$ 6 T. It resembles a Schottky anomaly
and is apparently related to the Zeeman effect of the Gd$^{3+}$
moments in the high-field paramagnetic state. All these phenomena
are essentially consistent with the results reported
previously,\cite{Champion2, Ramirez, Petrenko1} which were
discussed to originate from the partial lift of degenerate
states.\cite{Ramirez} The specific-heat data in $H \parallel$
[111] show rather similar behaviors to those in $H \parallel$
[110], but with some obvious differences in the positions and
magnitudes of those peaks, which indicates a non-negligible
magnetic anisotropy of Gd$_2$Ti$_2$O$_7$.\cite{Hassan, Petrenko1,
Glazkov2}

\begin{figure*}
\includegraphics[clip,width=14.5cm]{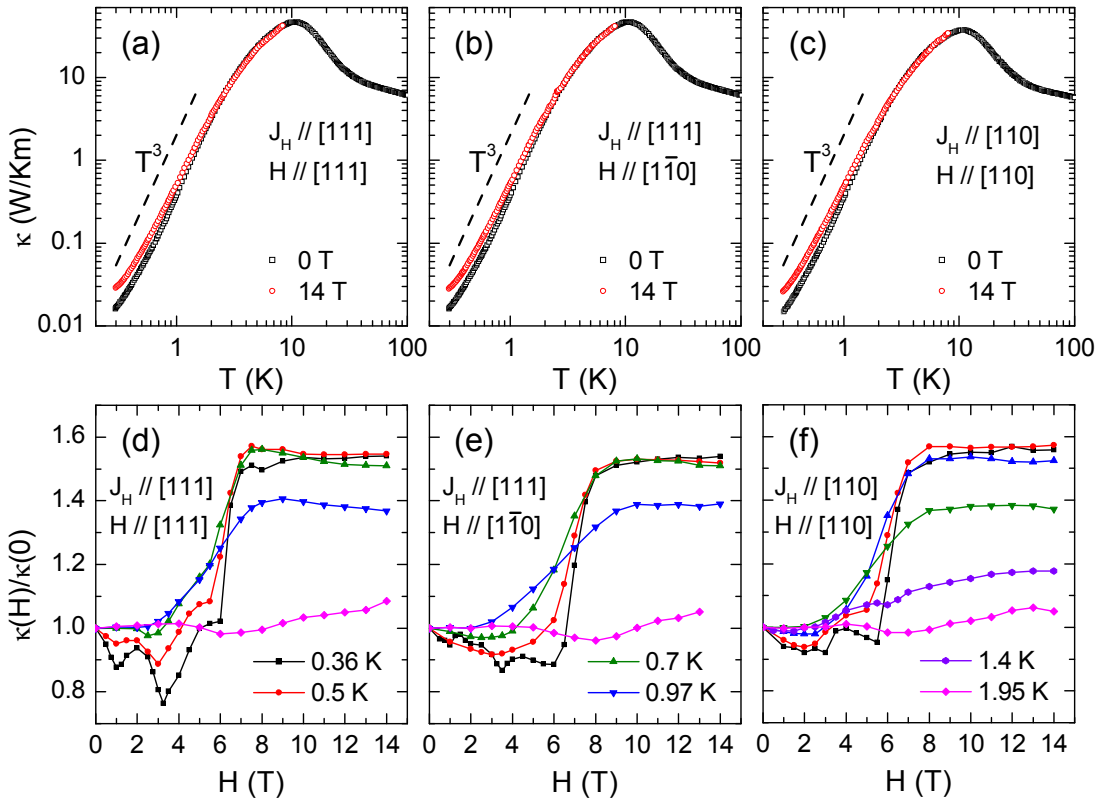}
\caption{(Color online) (a-c) Temperature dependencies of thermal
conductivities of Gd$_2$Ti$_2$O$_7$ single crystals in zero and 14
T. The directions of heat current ($J_H$) and magnetic field are
along either the [111] or the [110] directions. Note that in panel
(b), the magnetic fields along the [1$\bar{1}$0] direction
(equivalent to the [110]) are actually perpendicular to the [111]
direction. The dashed lines indicate a $T^3$ temperature
dependence. (d-f) The corresponding magnetic-field dependencies of
thermal conductivities at low temperatures for the three different
configurations. Two sets of data for $J_H
\parallel$ [111] were taken on a sample having a size of
3.9$\times$0.63$\times$0.17 mm$^{3}$. Another sample for $J_H
\parallel$ [110] has a size of 2.4$\times$0.65$\times$0.16
mm$^{3}$. The heat currents were flowing along the longest
dimensions of these samples.}
\end{figure*}

Figures 2(a)-2(c) show the thermal conductivity as a function of
temperature in zero and 14 T fields for different directions of
magnetic field and heat current ($J_H$). Apparently, the heat
transport of Gd$_2$Ti$_2$O$_7$ is almost isotropic in both zero
and 14 T fields and is independent of the field direction. The
zero-field $\kappa(T)$ curves exhibit a typical phonon transport
behavior at high temperatures and large phonon peaks at about 10
K.\cite{Berman} It is notable that at the phase transitions from
the high-$T$ paramagnetic state to the low-$T$ magnetically
ordered state at 1.0 and 0.7 K, indicated by the specific-heat
data, there is no obvious change of $\kappa(T)$. It seems that the
magnetic excitations of Gd$_2$Ti$_2$O$_7$ have no obvious
contribution to transporting heat. However, at low temperatures
the $\kappa$ is increased in 14 T, indicating that there is
magnetic scattering on phonons in zero field that can be removed
by applying high field.\cite{Zhao_GFO}

The detailed magnetic-field dependencies of the low-$T$ thermal
conductivities are shown in Figs. 2(d)-2(f). For three different
configurations, $\kappa(H)$ isotherms show very similar behaviors.
At very low temperatures, the $\kappa$ shows some reduction at low
fields, a step-like increase at about 6 T, and a high-field
plateau. Upon increasing temperature, the low-field reduction is
weakened and the transition at 6 T is broadened, accompanied with
a decrease of the high-field plateau. At 1.95 K, the $\kappa$ is
almost independent of field. At lowest temperature, the high-field
plateau is started from 7--8 T. From the specific-heat data, it is
known that above 8 T the samples enter the paramagnetic or the
high-field polarized state. Therefore, the step-like increase and
high-field plateau of $\kappa$ are apparently originated from the
quench of the phonon scattering by magnetic excitations (magnons
from the ordered state), when entering the spin-polarized
state.\cite{Zhao_GFO} This means that the low-$T$ thermal
conductivity in high fields would be a phonon transport free from
magnetic scattering. However, one may note that at sub-Kelvin
temperatures the 14-T $\kappa(T)$ curves have some obvious
curvatures in the log-log plot instead of showing a power law
close to $T^3$. It is deviated from the expectation of a phonon
heat transport in the boundary scattering limit.\cite{Berman} This
unusual temperature dependence points to some peculiarity of the
high-field magnetic state of Gd$_2$Ti$_2$O$_7$. Nevertheless, it
is rather clear that magnons in this material mainly play a role
of scattering phonons and there is no signature that they can
contribute substantially to transporting heat.

It can also be seen that the low-field behaviors of $\kappa(H)$
are a bit different for three different configurations of heat
current vs field direction. For $H \parallel$ [111], as shown in
Fig. 2(d), the 0.36-K $\kappa(H)$ curve has two pronounced ``dips"
at 1 and 3.25 T. With increasing temperature, they become more
shallow and disappear above 1 K. For $H \parallel$ [110] (or
$\perp [111]$), as shown in Figs. 2(e) and 2(f), the 0.36-K
$\kappa(H)$ curves show two weak ``dips", which are slightly
different between $J_H \parallel$ [111] and $J_H \parallel$ [110].
In the former case, they locate at about 1 and 3.5 T, while in the
latter case they locate at about 1 and 3 T (the 1 T dip in Fig.
2(f) is very weak). Note that in these cases, the relative
directions between the heat current and magnetic field are
different, which yields some difference in the demagnetization
effect. Considering this factor, the positions of two dips depend
on the direction of magnetic field rather than that of heat
current. These two dips disappear quickly with increasing
temperature.

It should be noted that these low-$T$ $\kappa(H)$ isotherms do not
display obvious hysteresis, which indicates that the possible
magnetic domains\cite{Petrenko2} are not coupled with the phonon
transport. In this regard, the $\kappa(H)$ results are not able to
distinguish the order of phase transition at 6 T, which either a
weak first-order transition or a second-order one were suggested
by some earlier experimental results.\cite{Petrenko1, Bonville}

\begin{figure}
\includegraphics[clip,width=6cm]{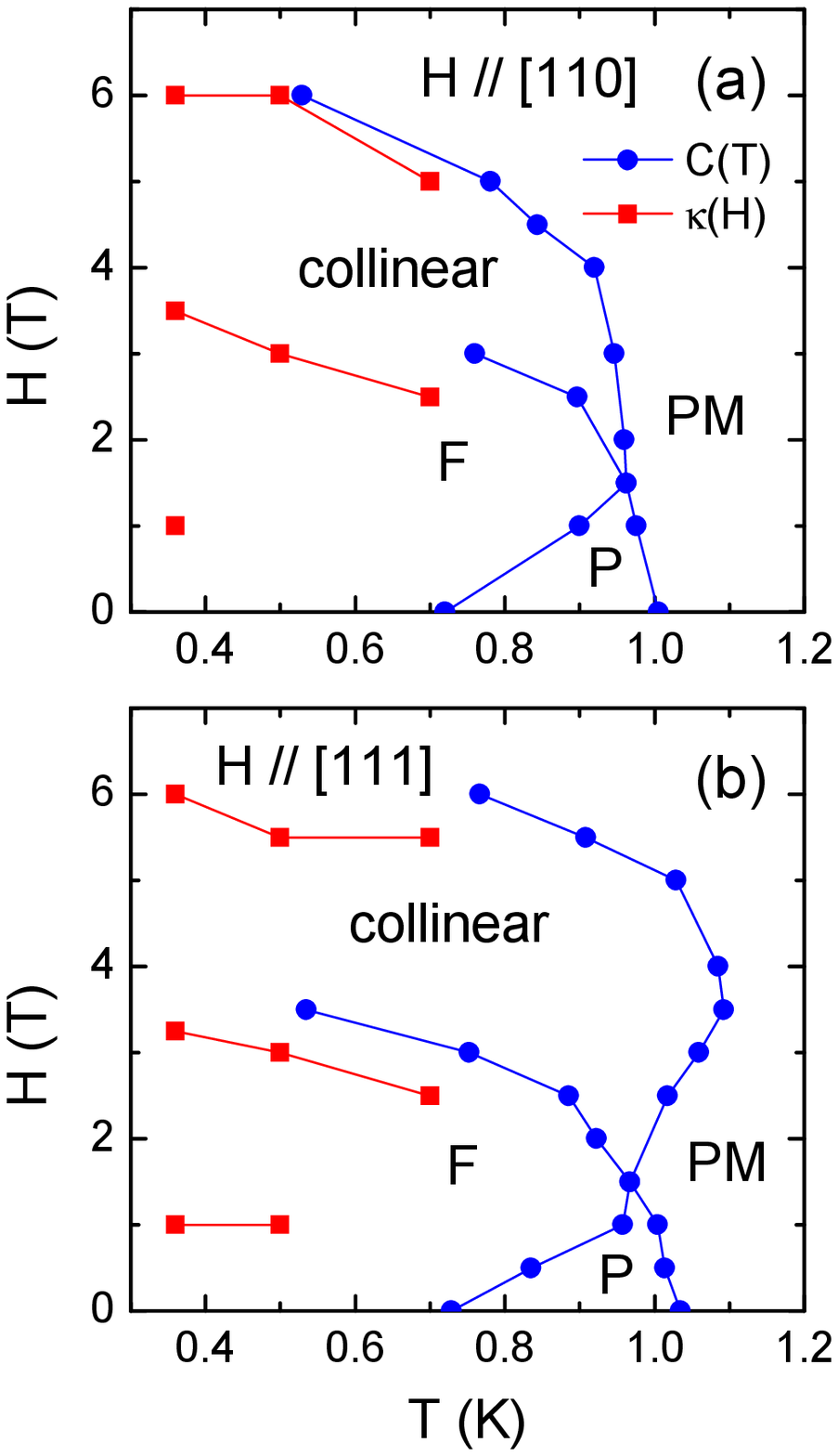}
\caption{(Color online) Temperature-field phase diagrams of
Gd$_2$Ti$_2$O$_7$ single crystals obtained from the specific-heat
data for $H \parallel$ [110] and $H \parallel$ [111]. The
characteristic fields of the $\kappa(H)$ curves are also plotted
for the comparison. $PM$ is the high-temperature paramagnetic
state. The $P$, $F$, and collinear states are three different
phases mentioned in the main text.}
\end{figure}

To probe the origin of the $\kappa(H)$ behaviors, we compare the
characteristic fields, including the dips and step-like increase,
with those transitions from the specific-heat data. In Fig. 3, the
$H-T$ phase diagrams of Gd$_2$Ti$_2$O$_7$ for either $H \parallel$
[111] or $H \parallel$ [110] are determined by the transitions
from the specific heat. They are nearly the same as those in some
earlier literature.\cite{Petrenko1} It is known that there are at
least four different magnetic phases in the phase diagram, that
is, the high-$T$ and high-field paramagnetic (PM) phase, the
low-$T$ and low-field $F$ state, a lower-field $P$ state, and a
higher-field collinear state (see the Introduction section). It is
found that the characteristic fields from the $\kappa(H)$ data
have a good correspondence with the phase boundaries among these
four phases. The step-like increase of $\kappa$ at about 6 T is
caused by the transition from the low-field ordered state to the
paramagnetic state.\cite{Zhao_GFO} It can be explained as the
elimination of the magnon scattering on phonons in a
spin-polarized state. The dip of $\kappa(H)$ at about 3 T is
caused by the transition from the $F$ state to the collinear
state. It seems that this transition is a rearrangement of the
spin structure, which is usually associated with the collapse of
the spin anisotropy gap and the increased numbers of low-energy
magnons. The phonon scattering is enhanced by a large number of
magnons at the transition fields.\cite{Zhao_GFO, Wang_HMO,
Wang_TMO} Another dip of $\kappa$ at about 1 T, however, has no
direct relationship to the specific-heat data. In the phase
diagrams determined by the specific heat, this transition is
locating deeply in the $F$ state. In this regard, it may be
related to a very recent experimental finding of a new phase
boundary at 1 T for $H \parallel$ [100].\cite{Petrenko2} This
boundary separates some unknown phases (named as $I$ and $I'$ in
Ref. \onlinecite{Petrenko2}) from the $F$ state. Since the phase
diagrams of Gd$_2$Ti$_2$O$_7$ are rather similar for different
field directions,\cite{Petrenko1} it is reasonable to believe that
these newly found phases and phase boundaries exist also in the
cases of $H \parallel$ [111] and $H \parallel$ [110]. Finally, it
should be pointed out that the small differences in the critical
fields between the $\kappa(H)$ data and the specific-heat data are
probably due to either the uncertainty in defining the transition
fields in two different physical properties or the difference in
the demagnetization effect, or both.

\subsection{Er$_2$Ti$_2$O$_7$}

\begin{figure}
\includegraphics[clip,width=6.0cm]{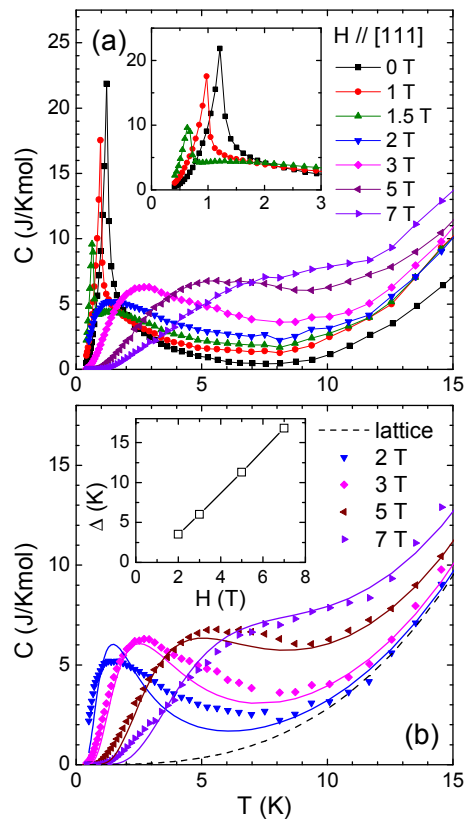}
\caption{(Color online) (a) Temperature dependencies of the
specific heat of Er$_2$Ti$_2$O$_7$ single crystal for magnetic
field along the [111] axis. Inset: Zoom-in of the low-temperature
plot. The sample size is 0.46$\times$0.67$\times$0.08 mm$^{3}$,
with the field direction along the shortest dimension. (b) The
fits of the high-field data ($H \ge$ 2 T) to Eq. (\ref{SH}) as
described in the main text. The solid lines represents the fitting
results and the dashed line shows the phonon contribution. Inset
displays the fitting parameter $\Delta$.}
\end{figure}

Figure 4(a) shows the low-$T$ specific heat of Er$_2$Ti$_2$O$_7$
single crystal. In zero field, a sharp $\lambda$ peak appears at
about 1.2 K, which reveals a second-order phase transition from
the paramagnetic state to the long-range ordered state. Applying
magnetic field along the [111] direction, the peak is gradually
suppressed and shifts to lower temperatures. It is not observable
at temperatures down to 0.4 K with $H \ge$ 2 T. At the same time,
a much broader peak emerges at about 1.3 K when $H =$ 1.5 T. It
shifts to higher temperatures and becomes even broader with
increasing field, which behaves like a Schottky
anomaly.\cite{Ruff, Sosin2} All these are in good agreement with
the previous results.\cite{Champion1, Ruff, Sosin2} In
literatures, the specific heat of Er$_2$Ti$_2$O$_7$ were studied
for three different directions of magnetic field, that is, $H
\parallel$ [100], [110], and [111], and were found to show some
anisotropy of the transition fields between the ground state and
the high-field quantum paramagnetic state. Furthermore, the
high-field specific heat were dependent on the field direction. In
particular, the high-field humps or broad peaks in the [100] and
[110] fields could be described by a simple two-level Schottky
anomaly, while those in the [111] fields could not.\cite{Sosin2}
Our present data confirmed this. As shown in Fig. 4(b), the
high-field data ($H \ge$ 2 T) are fitted using the following
formula

\begin{equation}\label{1}
C = \alpha R
(\frac{\Delta}{k_BT})^{2}\frac{e^{-\Delta/k_BT}}{(1+e^{-\Delta/k_BT})^{2}
}+ \beta T^3, \label{SH}
\end{equation}
where the first term represents the two-level Schottky anomaly and
the second one is the phonon contribution. Here, $R$ is the
universal gas constant, $\alpha$ is a numerical coefficient,
$\Delta$ is the gap value and is dependent on the Zeeman effect,
and $\beta$ is the coefficient for low-$T$ phonon specific heat.
It is found that this formula could qualitatively describe the
data but the fittings are not perfect, which is very similar to
the earlier study by Sosin {\it et al.}\cite{Sosin2} Note that the
parameters in our fittings, $\alpha$ = 1.65 and $\beta$ = 2.8
$\times 10^{-3}$ J/K$^4$mol, are also almost the same as those in
Ref. \onlinecite{Sosin2}. All these indicate that for $H
\parallel$ [111], higher-field Schottky-like peaks are not a simple
behavior of paramagnetic moments. Some soft modes related to the
spin fluctuations are probably involved.\cite{Ruff, Sosin2}

In a theoretical scenario, Er$_2$Ti$_2$O$_7$ was found to have a
ground state of quantum order by disorder and exhibit a quantum
critical point at 1.5--2 T, where the long-range order is
suppressed.\cite{Zhitomirsky, Savary, Sosin2} However, the
zero-field long-range order actually coexists with spin
fluctuations, as revealed by the magnetic resonance spectroscopy
and inelastic neutron scattering.\cite{Ruff, Sosin2} These soft
spin excitations would not only contribute to the specific heat
but also be easily coupled with low-energy phonons and
significantly damp the phonon transport.\cite{Li_TTO} Across the
quantum critical field (1.5--2 T), the magnetic excitations of
ground state change drastically, which in principle could be
detected by the thermal conductivity.

\begin{figure*}
\includegraphics[clip,width=14.5cm]{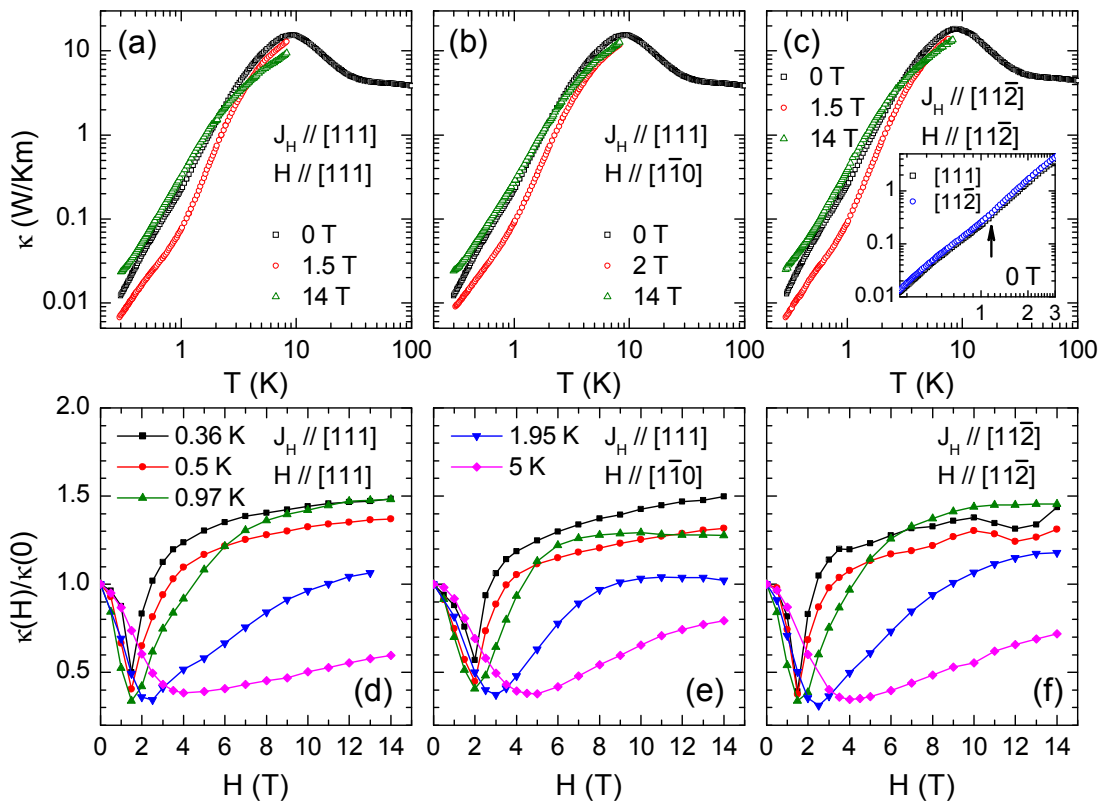}
\caption{(Color online) (a-c) Temperature dependencies of thermal
conductivities of Er$_2$Ti$_2$O$_7$ single crystals for different
directions of heat current ($J_H$) and magnetic field. In panel
(b), the magnetic fields along the [1$\bar{1}$0] (equivalent to
the [110]) are actually perpendicular to the [111] direction. In
panel (c), both the heat current and the magnetic fields are along
the [11$\bar{2}$] direction, which is actually perpendicular to
the [111] axis. The inset to panel (c) shows the zoom in of the
zero-field $\kappa(T)$ curves in temperature regime of 0.3--3 K,
and the arrow indicates a weak change in slopes of these data.
(d-f) The corresponding magnetic-field dependencies of thermal
conductivities at low temperatures for the three different
configurations. The sample sizes are 3.4$\times$0.65$\times$0.17
mm$^{3}$ for $J_H \parallel$ [111] and 4.0$\times$0.65$\times$0.18
mm$^{3}$ for $J_H \perp$ [111], respectively. The heat currents
were flowing along the longest dimensions of them.}
\end{figure*}

Figure 5 shows the temperature and magnetic-field dependencies of
the thermal conductivity for different directions of heat current
and magnetic field. Both the $\kappa(T)$ curves and the low-$T$
$\kappa(H)$ isotherms show no obvious anisotropy. As far as the
zero-field $\kappa(T)$ data are concerned, the phonon peaks locate
at about $8$ K; below about 1.2 K, as shown in the inset to Fig.
5(c), some slight changes in the slopes of these $\kappa(T)$ can
be seen, which is related to the phase transition from the
paramagnetic state to the long-range ordered state. With applying
field of 1.5 or 2 T, the $\kappa$ become much smaller than the
zero-field data and the changes of the $\kappa(T)$ slopes at
$\sim$ 1.2 K are more significant. In a high field of 14 T, the
$\kappa$ values are larger than the zero-field values at low
temperatures, which indicates that in the spin-polarized state,
the phonon scattering by magnetic excitations is quenched and the
phonon transport is recovered. However, at relatively higher
temperatures, the $\kappa$ in 14 T field is much smaller than the
zero-field value, suggesting a field-induced magnetic scattering
effect. This is different from the high-$T$ behavior of
Gd$_2$Ti$_2$O$_7$.

The $\kappa(H)$ isotherms show a dip-like behavior at low fields
and an increase at high fields, particularly at very low
temperatures. At the first glance, it is similar to that caused by
paramagnetic scattering on phonons.\cite{Li_NGSO, Sun_PLCO,
Sun_GBCO} But one may note a remarkable difference between the
Er$_2$Ti$_2$O$_7$ results and the paramagnetic scattering
phenomenon. In the latter case, the minimum or dip of the
$\kappa(H)$ isotherm is due to a phonon resonant scattering by
some magnetic excitations between the energy levels that are
affected by the Zeeman effect;\cite{Li_NGSO, Sun_PLCO, Sun_GBCO,
Berman} therefore, the position of the dip should shift almost
linearly with the magnetic field. For Er$_2$Ti$_2$O$_7$, at higher
temperatures ($>$ 1 K), the dip of $\kappa$ becomes broader and
shifts to higher fields with increasing temperature, which is
indeed the same as the paramagnetic scattering effect. However, as
can be seen in Figs. 5(d)-5(f), the dip fields are temperature
independent for $T <$ 1 K. Apparently, this sub-Kelvin behavior
must have some other origin. Since the low-$T$ dip fields (1.5 or
2 T) are close to the characteristic field in the specific heat,
above which the $\lambda$ peak was suppressed, it is likely that
they correspond to the phase transition from the low-field ordered
state to the high-field quantum paramagnetic state. In passing, it
should be pointed out that the dip field is essentially isotropic
for different field directions, as shown by Figs. 5(d) and 5(f),
while the higher dip fields in Fig. 5(e) could be due to the
demagnetization effect. In addition, the $\kappa(H)$ curves at
0.36 and 0.5 K in Fig. 5(f) display shallow minimums at high field
($>$ 10 T). It is unclear whether they are related to some kind of
transition since the feature is rather weak and broad.

\begin{figure}
\includegraphics[clip,width=7cm]{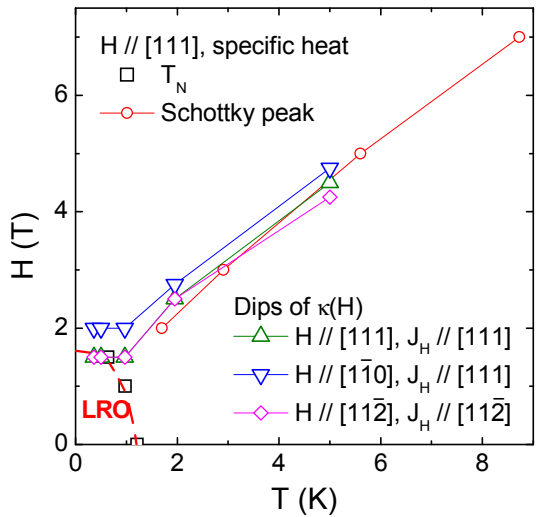}
\caption{(Color online) Temperature-field phase diagram of
Er$_2$Ti$_2$O$_7$ single crystal for $H \parallel$ [111] from the
specific-heat data. The low-$T$ $\lambda$ peak gives the
antiferromagnetic transition temperature $T_N$. The field
dependence of $T_N$ displays the phase boundary of the long-range
ordered (LRO) ground state, as shown by the dashed line. The
higher-field Schottky-like peaks may separate the ``high-$T$,
low-field" thermally paramagnetic state and the ``high-field,
low-$T$" quantum paramagnetic state, as suggested by Ref.
\onlinecite{Ruff}. The characteristic dip fields of the
$\kappa(H)$ isotherms are also shown for three configurations of
the heat current ($J_H$) and magnetic field.}
\end{figure}

Figure 6 shows the comparison of dip fields of the $\kappa(H)$
with the characteristic fields of the specific heat. It is clear
that the dips of $\kappa(H)$ have no good correspondence with the
phase boundary of the AF state, and at different temperature
regions they have very different origins. At very low temperatures
($T <$ 1 K), the dips are coincided with the quantum critical
field that suppresses the long-range-ordered ground state, which
can be ascribed to be a simple explanation of the strong phonon
scattering by the critical magnetic fluctuations. This is
supported by a complete mode softening of spin excitations at
$H_c$ as revealed by the inelastic neutron scattering.\cite{Ruff}
The higher-field enhancement of $\kappa$ is due to the removal of
low-lying magnetic excitations at $H > H_c$.\cite{Ruff} On the
other hand, the higher-$T$ ($T >$ 1 K) dips have a good
correspondence with the higher-field Schottky-like anomalies of
the specific heat. Although this anomaly was discussed to separate
the ``high-$T$, low-field" thermally paramagnetic state and the
``high-field, low-$T$" quantum paramagnetic state,\cite{Ruff} the
paramagnetic excitations are the main contribution and are likely
playing a role of scattering phonon. Therefore, it is
understandable that the $\kappa(H)$ behaviors at $T >$ 1 K match
well with the phonon resonant scattering effect by the
paramagnetic ions.

\section{DISCUSSION}

Both Gd$_2$Ti$_2$O$_7$ and Er$_2$Ti$_2$O$_7$ display the magnetic
excitations scattering on phonons. It seems to be a common
phenomenon in pyrochlore $R_2$Ti$_2$O$_7$. In spin-ice members
Dy$_2$Ti$_2$O$_7$ and Ho$_2$Ti$_2$O$_7$, there is negligible
scattering on phonons by the magnetic excitations in zero
field.\cite{Fan_DTO, Kolland, Toews} This is due to the fact that
the peculiar magnetic excitations, magnetic monopoles, in the
spin-ice compounds have sizeable energy gaps that prevent the
thermal activations at very low temperatures.\cite{Jaubert}
However, in magnetic fields the magnetic monopoles can be well
populated and were found to scatter phonons rather strongly,
particularly at the field-induced magnetic transitions between the
``spin-ice", the ``kagom\'e-ice", and the ``three-in-one-out"
states, although they might be able to transport heat
also.\cite{Fan_DTO, Kolland, Toews} In the quantum-spin-liquid
material Tb$_2$Ti$_2$O$_7$, the coupling between spin fluctuations
and phonons is so strong that the phonon heat transport exhibits a
glassy-like behavior.\cite{Li_TTO} In comparison, the coupling or
scattering effect between magnetic excitations and phonons in the
long-range ordered Gd$_2$Ti$_2$O$_7$ and Er$_2$Ti$_2$O$_7$ are not
very strong but detectable at zero field.

One reason that the magnetic excitations act mainly as phonon
scatterers rather than heat carriers in $R_2$Ti$_2$O$_7$ is
apparently related to the rather weak exchange interactions in
these materials. As a result, the magnetic excitations have weak
dispersion or small velocity. An example is that the magnetic
monopoles in the spin-ice compounds were estimated to have a small
velocity of $\sim$ 20 m/s.\cite{Kolland} The magnon velocity of
Er$_2$Ti$_2$O$_7$, calculated from the dispersion curve, is about
80 m/s only.\cite{Dalmas} Gd$_2$Ti$_2$O$_7$ is expected to have a
similar situation, since the exchange energies of these materials
are essentially comparable.\cite{Raju, Dunsiger, Cao1, Dalmas} In
contrast, the magnon velocity is about 2--4 orders of magnitude
larger in many materials that show substantial magnon heat
transport.\cite{Hess1, Hess2, Sologubenko}

Although the heat transport of Gd$_2$Ti$_2$O$_7$ and
Er$_2$Ti$_2$O$_7$ have some similarities in the weak anisotropy
and magnetic scattering effect, there are several notable
differences between them. First, Gd$_2$Ti$_2$O$_7$ shows much
larger phonon peaks in the zero-field $\kappa(T)$ curves, even
though these two compounds have the same crystal structures. For
nonmagnetic insulators, larger phonon peak may indicate less
structural imperfections in crystals. However, for magnetic
materials, the magnetic scattering can play a role. This
possibility seems to be supported by the details of the
temperature dependence of $\kappa$. As can be seen in Figs. 2 and
5, near the long-range order transition in zero field, the
$\kappa(T)$ of Gd$_2$Ti$_2$O$_7$ do not show any observable
changes, while those of Er$_2$Ti$_2$O$_7$ show weak changes of the
$\kappa(T)$ slopes. Second, the magnetic scatterings actually have
quite different origins in these two materials. In
Gd$_2$Ti$_2$O$_7$, the magnetic excitations of the long-range
ordered state, {\it i.e.}, magnons are effectively phonon
scatterers. For this reason, the magnetic-field-induced changes of
$\kappa$ are only presented at low temperatures; above 2 K, the
$\kappa$ is nearly unaffected by applying field. In
Er$_2$Ti$_2$O$_7$, however, the role of magnons in the long-range
ordered state is not so obvious. Nevertheless, in zero field and
at very low temperatures, magnons are likely to scatter phonons,
since the high-field $\kappa$ are about 50\% larger than the
zero-field values. On the other hand, the low-energy spin
excitations associated with the quantum critical point induce a
strong scattering on phonons in a rather broad temperature range.
Another contribution of the magnetic excitations is the high-$T$
paramagnetic scattering effect. In this regard, it is rather
difficult to understand why there is no any signature of
paramagnetic scattering effect in Gd$_2$Ti$_2$O$_7$. Third,
Gd$_2$Ti$_2$O$_7$ shows much more complicated $\kappa(H)$
behaviors than Er$_2$Ti$_2$O$_7$. This is clearly due to a more
complicated low-$T$ phase diagram of Gd$_2$Ti$_2$O$_7$. In present
work, the low-$T$ heat transport of these materials again shows a
close relationship to the magnetism and magnetic transitions, as
also found in many other magnetic materials.\cite{Fan_DTO,
Kolland, Sun_DTN, Zhao_GFO, Wang_HMO, Wang_TMO} It therefore
provides a useful tool to probe some unknown magnetic transitions,
as shown in Fig. 3.

There are also some details remained to be further studied. As
shown in Fig. 2, the temperature dependencies of $\kappa$ of
Gd$_2$Ti$_2$O$_7$ show some clear deviations from the power low at
sub-Kelvin temperatures even in 14 T field. This may suggest that
the magnetic scattering in the high-field spin-polarized state is
still active at such low temperatures. This unusual temperature
dependence points to some peculiarity of the high-field magnetic
state of Gd$_2$Ti$_2$O$_7$. One possible reason is that the ground
state of Gd$_2$Ti$_2$O$_7$ is a coexisting of long-range and
short-ranges orders. The short-range order or spin fluctuations
could survive in high field even when the long-range AF order was
suppressed. A supportive experimental result is that the magnetic
resonance spectroscopy has detected some weakly dispersive soft
modes in the spin-saturated phase, which could be explained based
on the spin-wave calculations.\cite{Sosin1}

\section{SUMMARY}

Thermal conductivities of pyrochlore $R_2$Ti$_2$O$_7$ ($R$ = Gd
and Er) are studied at low temperatures down to 0.3 K and in high
fields up to 14 T. It is found that the magnetic excitations play
a role of scattering phonons rather than transporting heat,
although these two materials have long-range magnetic orders at
low temperatures. In Gd$_2$Ti$_2$O$_7$, the field-induced magnetic
transitions can cause drastic changes of $\kappa$ at the phase
boundaries and lead to rather complicated field dependencies of
$\kappa$. The main phenomena include the dip-like features in
$\kappa(H)$ isotherms at the low-field phase transitions and a
step-like increase of $\kappa$ at the spin-polarization
transition. Er$_2$Ti$_2$O$_7$ shows much simpler field
dependencies of $\kappa$, with a dip-like feature caused by the
quantum phase transition at about 1.5--2 T or the paramagnetic
phonon scattering at higher temperatures. These data demonstrate
that heat transport is an effective means to probe the magnetic
properties of pyrochlore rare-earth titanates.

\begin{acknowledgments}

This work was supported by the National Natural Science Foundation
of China, the National Basic Research Program of China (Grants No.
2009CB929502 and No. 2011CBA00111), and the Fundamental Research
Funds for the Central Universities (Program No. WK2340000035).

\end{acknowledgments}

\end{document}